# The MC-QTAIM: A framework for extending the "atoms in molecules" analysis beyond purely electronic systems

*Shant Shahbazian*

*Department of Physics, Shahid Beheshti University, Tehran, Iran*
*E-mail: sh_shahbazian@sbu.ac.ir*

# Abstract

The quantum theory of atoms in molecules, QTAIM, is employed to identify AIM and quantify their interactions through the partitioning of molecule into atomic basins in the real space and it is confined only to the purely electronic systems composed of electrons as quantum particles and the nuclei as clamped point charges. The extended version of the QTAIM, called the multi-component QTAIM, MC-QTAIM, bypasses this border and makes it possible to identify AIM and quantify their interactions in systems composed of multiple quantum particles that electrons may or may not be one of their components opening a new door for the analysis of the exotic AIM and bonds. In this contribution, two conjectures, called Bader conjecture, BC, and extended Bader conjecture, EBC, are proposed as the cornerstones of the real-space partitioning of a molecule into atomic basins within the context of the QTAIM and the MC-QTAIM, respectively. A literature survey on various few-body quantum systems composed of quarks, nucleons, and elementary particles like muons and positrons is also done unraveling the fact that in all these diverse systems there are unambiguous cases of clusterizations. These clustered systems, irrespective to their components, behave as if they are molecules composed of some kind of atoms, instead of being an amorphous mixture of quantum particles. In the case of the muonic and the positronic molecules computational studies reveal that the AIM structures of these systems are well-captured by the EBC. Beyond identifying atomic basins, both QTAIM and MC-QTAIM attribute properties to AIM, which is their share from the molecular expectation values of quantum observables. It is demonstrated that not only the share from the average value of an observable may be attributed to an atomic basin, but also the fluctuation of each basin property is also quantifiable.

# Keywords

QTAIM, MC-QTAIM, hadronic molecules, nuclear molecules, exotic molecules

# I. Introduction

The idea that matter is composed of atoms, i.e. indivisible ingredients, goes far back in the history to the age of ancient Greeks, but laid dormant, though not completely forgotten, for more than two millennia.[1] In the modern era it was revived triumphantly through Dalton's atomic theory which attributes a distinct atom to each chemical element, and from then on the theory dominates the modern chemistry and physics.[2,3] Its central role in modern thinking is best described by Feynman that leaves no room for further comments: "If, in some cataclysm, all of scientific knowledge were to be destroyed, and only one sentence passed on to the next generations of creatures, what statement would contain the most information in the fewest words? I believe it is the atomic hypothesis (or the atomic fact, or whatever you wish to call it) that all things are made of atoms—little particles that move around in perpetual motion, attracting each other when they are a little distance apart, but repelling upon being squeezed into one another. In that one sentence, you will see, there is an enormous amount of information about the world, if just a little imagination and thinking are applied".[4]

With such a respectable history, it is probably shocking to a non-expert observer that the concept of "atom in a molecule" is a controversial topic subject for theoretical community and a source of disputes and discussions,[5,6] as well as a room for creative theoretical work.[7,8] The main problem with the concept of atom in a molecule is simply the fact that "Daltonian" atom is currently not anymore the atom conceived by Leucippus and Democritus; its status as the "indivisible" was lost to the elementary particles.[9] In fact, to current chemists and condensed matter physicists, the *true atoms are electrons and nuclei*. The nuclei are not literally indivisible but beyond nuclear physics and chemistry, for almost all other practical purposes, it is safe to ignore their internal structure and may be treated as if they are "effectively" elementary. In that sense, at the most fundamental level a molecule or a crystal is conceived to be composed of electrons and nuclei and the formulation of basic quantum equations is based on their properties and interactions. Accordingly, as emphasized masterfully by Laughlin and Pines,[10] the "theory of everything" for almost all of chemistry and condensed matter physics is Schrödinger's equation governing electrons and nuclei. This triggers a dichotomy in the above-mentioned disciplines since in the electronic Schrödinger equation there is no mention of Daltonian atoms. Whereas, the phenomenological models and the general language of these disciplines are full of the jargon of the Daltonian atomic theory. For a chemist, water molecule is composed of two hydrogen and one oxygen atoms while for a condensed matter physicist, the crystalline salt is composed of sodium and chlorine ions. In this jargon, atoms bear electric charges and dipoles, they interact and may bond to each other; in other words, they are seemingly well-characterized entities with properties and interaction modes. Phenomenological models of molecules and condensed phases like molecular mechanics and molecular dynamics are computational incarnation of this viewpoint.[11,12]

To remedy this dichotomy, and to reconcile the language of the "theory of everything" with the language of the Daltonian atomic theory, many researchers have tried to suggest methodologies to "extract" atoms in molecules (AIM) from the solutions of the electronic Schrödinger equation.[13–16] The basic premise of these methodologies is that AIM and their properties are somehow "buried" in the electronic wavefunctions,[16] reduced densities matrices,[15] electron densities,[13,14] and if these are "mined" properly, AIM can be retrieved. Since the basic principles of quantum mechanics are silent on how this mining should be realized, the proposed methodologies are inherently "heuristic" and this is the origin of the controversies and disputes around the best way of introducing the concept of AIM. A typical response to this problem is dismissing the whole issue as a "pseudo-problem" and assuming that the concept of AIM while useful, to be intrinsically vague and not amenable to a rigorous theoretical analysis within quantum mechanics.[5] Indeed, there are other seemingly vague concepts that despite their usefulness, may defy a rigorous analysis and their omission could undermine the basic structures of modern scientific disciplines. "Species" in taxonomy and evolutionary biology and "consciousness" in neuroscience and psychology are illustrative exmaples.[17,18] However, as discussed elsewhere,[6] it is hard, if even possible in principle, to demonstrate the intrinsic vagueness of a concept conclusively.[19] There are indeed historical examples where a seemingly vague concept has been transformed into a rigorous entity through introduction of a novel theoretical framework where the "affinity" may serve as a classic example. The concept of affinity in chemistry has a tortious history and has been a vague concept for centuries.[20] Finally, it was placed on a rigorous basis by de Donder through employing then newly proposed concept of the chemical potential within modern thermodyanmics.[21] Hopefully, the final fate of the concept of AIM is also yet to be determined but in the meantime, the mentioned heuristic methodologies may serve as temporary theoretical frameworks to analyze the properties and interaction modes of AIM as far as possible.

For more than a decade, present author and his coworkers were particularly focused on the framework of the quantum theory of atoms in molecules (QTAIM), which was championed by Bader and coworkers since the mid-seventies.[14,22–27] Our primary focus was on the mathematical foundations of the QTAIM itself.[28,29] Eventually, the research programme shifted to generalization of the methodology to be applicable to the "multi-component" quantum systems, which are molecular systems containing other quantum particles apart from electrons.[30–37] The central question was to check whether a generalized version of the QTAIM may reveal AIM in molecular systems where electrons are not the sole quantum particles. Although the origins of this problem has been reviewed some time ago,[38] in the second section of this contribution a brief survey is done to demonstrate why this is a legitimate question. Interestingly, contemplating various physical systems as composed of real or effective elementary particles, and at the same time perceiving them as composed of Daltonian-type atoms is far more widespread than usually conceived by chemists.[39] Some examples will be discussed in the second section.

Our generalized methodology, called the multi-component QTAIM (MC-QTAIM), was applied to ab initio wavefunctions of various so-called "exotic" molecular species composed of electrons and other elementary particles like positrons and muons revealing the underlying AIM structure in these species.[40–48] Novel concepts and features emerged from these studies including the regional positron affinities,[41] the positronic bond,[48] and the positive/negative muon's capability/incapability to shape its own atomic basin.[43,45,46] Also, when applied to the purely electronic systems, the MC-QTAIM analysis recovers the results of the QTAIM analysis, demonstrating that the former encompasses the latter just as a special case.[32]

In the second and third sections of this chapter, a comparative analysis is also done on the formulation of the QTAIM and the MC-QTAIM and the basic ideas behind these methodologies are articulated. In this articulation a new theoretical ingredient of the QTAIM, i.e. property fluctuation of AIM, is introduced which goes beyond the well-established formalism of the usual particle number fluctuation of AIM.[14,49,50] At the next step this new ingredient will be extended within the context of the MC-QTAIM. Particularly, the original idea of Bader that AIM are somehow "open" entities amenable to particle and property exchange,[14] is quantified by this new ingredient in a firm manner. Finally, in the fourth section some future opportunities for theoretical developments of the MC-QTAIM and even beyond are briefly discussed.

## II. Revealing AIM beyond purely electronic systems

### II.A. AIM as conceived within the QTAIM

Within context of the QTAIM, each atom in a molecule is conceived as a 3-dimensional basin in real-space enclosed by 2-dimensionl boundaries.[14] These boundaries are the zero-flux surfaces emerging from the local zero-flux equation of the gradient of the electron density: $\vec{\nabla}\rho(\vec{r}).\vec{n}(\vec{r})=0$.[14] Note that: $\rho(\vec{r})=N\int d\tau' \Psi^*\Psi$, where $\Psi$ is the electronic wavefunction of a purely electronic system and $N$ is the number of electrons while $d\tau'$ implies summing over spin variables of all electrons and integrating over spatial coordinates of all electrons except one arbitrary electron and $\vec{n}(\vec{r})$ is the unit vector normal to the surface. Bader called these basins interchangeably atomic basins, AIM, topological atoms, or quantum atoms,[14] but in this chapter they are called atomic basins. In a recent review paper the origin of these basins has been scrutinized; thus, the details are not reiterated herein and only the main points are reemphasized.[29]

Let us start from the fact that molecular electron densities are to a good approximation the sum of electron densities of the constituent atoms, and the atomic densities, i.e. electron density of free atoms, deform only marginally in molecules.[51–53] Particularly, the main trait of an atomic density, namely its maximum at the nucleus and "monotonic" decaying away from nucleus,[54,55] retain within molecules. This "robustness" of the atomic densities within molecular environment guarantees that molecular densities usually have a simple topography around equilibrium geometries, containing local maximum at each nucleus. In the language of topological analysis of electron densities,[14] a (3, -3) critical point (CP) is located at each nucleus and all zero-flux surfaces that do not cross these CPs are the boundaries of AIM, partitioning the real-space exhaustively into atomic basins. All gradient paths within an atomic basin, originating at infinity or from other types of CPs on its boundaries,[14] are ultimately terminating at (3, -3) CPs. Thus, an atomic basin is composed of a (3, -3) CP and its basin of "attraction" and the nucleus and all the electronic population within the basin of attraction belong to that basin. Whether these basins are appropriate representatives of Daltonian atoms is in principle a disputable matter but within the context of the QTAIM, this equivalence is an axiom. Interestingly, not the monotonicity nor the robustness of atomic densities have been proven rigorously from the first principles of quantum mechanics yet,[56–62] and the following "conjecture" was proposed some time ago to justify the equivalence:[29]

*For a molecular system containing $N$ electrons, with position vectors $\{\vec{r}_i\}$ and inter-electronic distances $\{r_{ij} = |\vec{r}_i - \vec{r}_j|\}$, and $Q$ clamped nuclei with atomic numbers $\{Z_\alpha\}$, with position vectors $\{\vec{R}_\alpha\}$ and inter-nuclear distances $\{R_{\alpha\beta} = |\vec{R}_\alpha - \vec{R}_\beta|\}$, described by the following electronic Hamiltonian in atomic units:*

$$\hat{H} = \left(-\frac{1}{2}\right)\sum_i^N \nabla_i^2 + \sum_i^N \sum_{j>i}^N \frac{1}{r_{ij}} + \hat{V}_{ext}, \qquad \hat{V}_{ext} = -\sum_i^N \sum_\alpha^Q \frac{Z_\alpha}{|\vec{R}_\alpha - \vec{r}_i|}$$

*There is always a critical distance between each pair of nuclei, denoted as $\{R_{\alpha\beta}^c\}$, that for geometries for which the inter-nuclear distances are larger than $\{R_{\alpha\beta}^c\}$, $\vec{\nabla}\rho(\vec{r};\{R_{\alpha\beta} > R_{\alpha\beta}^c\})$ derived from the ground electronic state contains just $Q$ numbers of (3, -3) CPs located at the position of the nuclei.*

Perhaps, this conjecture may rightfully be called the "Bader's conjecture" (BC); although it was not explicitly stated by him in the original literature of the QTAIM, it was tacitly assumed throughout developments of the methodology.[63] Indeed a wealth of computational studies on electron densities of numerous molecules and solids as well as experimental charge densities derived from the X-ray diffraction data point to the validity of the BC.[14,64–67] One may hope that someday BC will be proven and its status will be elevated to "Bader's theorem". By the way, let us briefly review some points regarding the current status of this conjecture.

Firstly, the role of $\hat{V}_{ext}$ is crucial in this conjecture since if one replaces Coulomb's potential with another arbitrary potential the whole conjecture may fall apart where a vivid example is the harmonic trap potential: $\hat{V}_{ext} = -\sum_{i}^{N}\sum_{\alpha}^{Q}\left(\omega_{\alpha}^{2}/2\right)\left(\vec{R}_{\alpha}-\vec{r}_{i}\right)^{2}$.[68] In this case Schrödinger's equation is analytically solvable and it is straightforward to demonstrate that for this particular external potential the number of (3, -3) CPs is irrelevant to the number of nuclei.[68] On the other hand, as discussed elsewhere,[69] it is possible to modify electron-nucleus Coulomb potential by taking into account the finite size of nucleus;[70] the electron density resulting from the short-range modified electron-nucleus potential yet conforms to the BC. Between these two extremes, there are various external potentials, e.g. magnetic interactions or relativistic corrections, which one may add to or replace with the Coulomb's point charge potential. Naturally, the question arises for which type of potentials the conjecture retains and at the best of author's knowledge, this is an open and untouched problem.[29,68] The reverse reasoning leads to: *for all external potentials fulfilling the BC the resulting basins are chemically well-defined regions within the context of the QTAIM since an atomic basin is attributed to each nucleus*. In fact, the BC reveals the basic principle behind the concept of AIM within the context of the QTAIM; *an atom in a molecule is a region with high "clustering" of electrons in the real-space*. In other words, as far as the real-space clustering is justified as a criterion to define Daltonian atoms, the BC may lead to a proper partitioning scheme for that system. This rational is also the basis of all subsequent discussions on the partitioning of systems beyond the purely electronic systems into atomic basins.

The second point is to stress that only the electron density of the electronic ground state, $\rho(\vec{r};\{R_{\alpha\beta}\})$, is the target of the conjecture, which stems from the fact that almost all original QTAIM computational studies were confined to these densities.[14] However, more recent studies by Bader himself,[71] as well as others,[72–76] point to the fact that at least for low-energy excited electronic states the local zero-flux equation is yet applicable and yields reasonable atomic basins. This may be reflected by encompassing low-energy excited electronic states into the conjecture, resolving the previously stated limitation to the ground state. However, since according to the best of author's knowledge, no systematic study has yet been conducted to check the validity of the conjecture for the whole spectrum of the excited electronic states of a molecule, this reasonable modification may be postponed. Indeed, the monotonicity of atomic electron densities is lost for the excited atomic states and thus it is very doubtful that the BC is applicable to high-energy excited molecular electronic states. This is because of the fact that the electron densities of high-energy excited states are exceedingly deformed in comparison to the ground-state electron density. One may conclude that most probably the BC is applicable to the ground and low-energy excited electronic states although before final settlement, further computational studies must be done to have a better perception on the true borderline between low- and high-energy electronic excited states. This point will be discussed again when considering possible applications of the MC-QTAIM to non-electronic systems.

The third point regarding BC is the "conditional" truth of the equivalence of the number of (3, -3) CPs and the number of nuclei for geometries for which the inter-nuclear distances are larger than a critical distance: $\{R_{\alpha\beta} > R_{\alpha\beta}^c\}$. This conditional truth stems from the fact that "non-nuclear" (3, -3) CPs, located in between nuclei, instead of being located at nuclei, appear if the inter-nuclear distance decreases sufficiently.[77–85] The associated basins are usually called "pseudo-atoms" and to be fair, this was noticed by Bader himself,[86] but it was usually treated more as an exception rather than the rule.[14] However, since the illuminating study of Pendas and coworkers it is known that for sufficiently close inter-nuclear distances pseudo-atoms may appear between the atoms of a system,[87,88] and even forming "quasi-molecules" when bonded to each other.[89,90] Even more, it was shown by Bader himself that for very close inter-nuclear distances the two atomic basins and the pseudo-atom in between merge and the two nuclei then reside in a single basin containing a single non-nuclear (3, -3) CP.[91] Such basin is usually called united-atom limit, once again first discovered by Bader and coworkers,[63] and curiously there are rare but illustrative examples that they may appear even at or around equilibrium geometries.[92] In contrast to these examples, it is fair to say that for most considered molecules at the ambient conditions the criterion: $\{R_{\alpha\beta}^{equilibrium} > R_{\alpha\beta}^c\}$ is satisfied, and this explains the general success of the local zero-flux equation in delineating reasonable atomic boundaries. However, the discussed counterexamples demonstrate that the electron clustering in the real-space does not have a "universal" pattern and one must be prepared to accept that the number of atomic basins varies sometimes considerably, upon large geometrical variations of the nuclei. Whether this may be interpreted as a weakness of the BC to recover

Daltonian atoms or conversely, a hint for extension of the original concept of Daltonian atom is a matter of taste and possibly dispute. Accordingly, in the extension of the QTAIM beyond purely electronic systems we need first to answer the basic question of how atomic basins must be defined in these systems. Nevertheless, let us first briefly review the evidence that demonstrates seeking Daltonian-type atoms beyond the purely electronic systems is a worthwhile enterprise.

### II.B. AIM beyond purely electronic systems

The previous discussion reveals an *intermediate level of organization* of the electronic matter, which manifests itself in the real-space clustering of electrons around nuclei resulting in new "composite" entities, i.e. AIM. In the case of electronic matter, the origin of this organization is the dominant interaction of the clamped nuclei and electrons. Accordingly, the question emerges whether there are other examples in nature when a similar organization emerges through other mechanisms manifesting itself by some type of real-space clustering. Interestingly, the answer is affirmative,[93] and herein some of better-known examples at the atomic and subatomic levels are considered briefly which are relevant to possible future extensions of the QTAIM. There are three main classes of systems, apart from the purely electronic matter, where quantum systems may organize themselves into a molecular-like structure and reveal a sort of intermediate level of organization. These classes include:

1-Hadronic molecules,

2-Nuclear molecules,

3-Exotic molecules,

While in this chapter only the technical details of the latter class is considered, a brief account of all these classes are given in this subsection.

At the most fundamental level, matter is composed of quarks and leptons and the hadrons are the group of composite particles made solely of quarks though this is by part a simplification since apart from quarks, gluons are also an important ingredient of hadrons.[9,94] Quarks have fractional electric charges, one-third and two-third of the electron's charge, and there are six "flavors" (or types) of them classified as three families, also sometimes called as generations (the usual stable matter around us is solely composed of the first/lightest of these generations).[9] The nature of forces responsible for hadrons' formation are quite distinct from the Coulomb forces acting within molecules and are based on a property called color charges of quarks, to be distinguished from the electric charges.[9] The forces between quarks are called color/strong forces and are described by the theory of quantum chromodynamics (QCD) (details of color charges and their relationship to the inter-quark interaction are not considered herein and may be found in relevant textbooks).[9,95] In fact, it is more appropriate to think of quark systems like the usual atomic systems;[96] an atom is composed of a handful of interacting elementary particles but has an infinite numbers of states, which are observable in the atomic spectroscopies. In the case of a quark system, i.e. a handful of interacting quarks, each state is conceived as a new hadron and because of this rationale, searching for and systematizing of hadrons are called hadron spectroscopy.[97,98] Originally, hadrons were assumed to be only two-quark (mesons) or three-quark (baryons) systems where the familiar ingredients of nuclei, proton and neutron, are examples of the latter group, while other hadrons are inherently unstable and decay to more stable particles/states.[9,95] However, more recently examples of tetraquark and pentaquark systems, also called exotic hadrons, have been discovered,[99–103] and the rules governing their construction is an active field of research.[104,105] A subset of these multiquark systems are called hadronic molecules since they are organized not just as an amorphous mixture of quarks but have sub-hadronic structures;[106] in simplest form this can be a di-hadronic structure for a tetraquark system.[103,107] Accordingly, a molecular tetraquark may be conceived as if it is composed of two tightly bonded mesons and it is indeed desirable to deduce this structure from ab initio QCD calculations. Hence, starting from quarks, their color properties and their interaction rules given by the QCD, various theoretical and computational methodologies have been utilized to deduce the molecular nature of the exotic hadrons.[108–114] Particularly relevant to our objective is ab initio lattice-QCD computational studies,[115–117] where the derived real-space color and energy densities are employed to visualize the real-space state of the normal mesons,[118,119] and the exotic hadrons.[120–126] The topographies of these densities are reminiscent of those of the electron densities but at the best of author's knowledge, no topological analysis has been conducted yet on these densities. We will not consider further this area in this contribution but it is tempting to seek for an extended BC-like conjecture to derive the "atomic" structure of the molecular hadrons, which is a completely uncharted territory, and worth exploring.

The next examples considered herein are nuclear molecules as a special subset of nuclei.[127–136] In principle, nuclei are also composed of quarks but at the usual densities encountered in nuclei, they are effectively composed of protons and neutrons, collectively called nucleons, bonded by strong nuclear forces. The details of strong forces between nucleons are generally more complicated than those operative between quarks since nucleons as hadrons, in contrast to quarks, do not carry color charges and are "colorless".[9,100] For comparison note that electrically charged particles within atoms are interacting through the universal Coulomb interactions. However, the atoms themselves are electrically neutral and their interactions, though also electromagnetic in nature, are described by "effective" force laws, which are indeed more complicated and less universal than the Coulomb interactions.[137] Thus, the ab initio "theory of everything" of nuclei is Schrödinger's equation, or a relativistic extension of it, governing the many-body system composed of nucleons interacting through effective strong forces.[138–143] Apart from the ab initio,[144–156] and the second-quantized based methods,[157,158] various models have been also proposed to unravel certain aspects or properties of nuclei.[159] Examples include the vibrational model,[160,161] the rotational model,[162] the shell model,[163,164] the interacting boson model,[165–168] and other collective models.[169,170] However for most nuclei, this many-body system is usually conceived as a droplet-like entity, usually called liquid-drop model of nucleus.[171] Within context of this model nucleus is conceived as an amorphous system with a spherical or deformed shape, however, there are a subset of nuclei that defies this picture and are called nuclear molecules. Accordingly, a new model of nuclei was proposed in 1930's where it was assumed that certain light nuclei carrying $4n$ nucleons ($n = 2,3,...$): $^8$Be, $^{12}$C, $^{16}$O, $^{20}$Ne, $^{24}$Mg, $^{28}$Si, …, are effectively composed of the alpha-particle (Helium-4 nucleus: $^4$He) clusters at their ground state.[172] Since then the $\alpha$ cluster model found to be applicable not only to the stable ground states of certain light nuclei, but also to various excited states formed during nuclear reactions,[173,174] $\alpha$-decay process,[175] and even for certain heavy nuclei.[176] Particularly, in contrast to the excited electronic states of molecules discussed previously, $\alpha$ clustering is even more pronounce for the excited nuclear states in the threshold of nuclear disintegration as it is usually systemized by the Ikeda diagram.[175] Of particular interest in the class of $\alpha$ clustering models is the close-packed spherical arranged geometrical model of $\alpha$ clustered nuclei, championed by Hafstad and Teller,[177] which was also pursued by Pauling.[178] Assuming a "bond" between each close-packed $\alpha$-$\alpha$ contact and attributing a single "bond energy" parameter to all $\alpha$-$\alpha$ bonds, this simple model is capable of reproducing the nuclear binding energy to a good accuracy.[177] The simplest quantum mechanical treatment of this geometrical model is to omit the nucleons altogether and treating each $\alpha$ cluster as a structureless point-particle interacting through the effective $\alpha$-$\alpha$ potentials.[179] However, in practice, this "coarse-grained" quantum model had a limited success and even for $^{12}$C, composed of just three $\alpha$ clusters,[177] it not easy to reproduce experimentally observed nuclear states and their properties without invoking three-body potential energy terms.[180,181] The need for three-body (or in general $n$-body) potential energy terms, which are not reducible to the usual two-body interaction potentials, is reminiscent of the genuine $n$-body inter-atomic potentials appearing in the theory of intermolecular forces.[137] This reveals the fact that the internal structure of $\alpha$ clusters is not intact within the nucleus and "internal deformations" are induced upon interactions with other $\alpha$ clusters. A natural step for improvement is to conceive the nucleus as if it is composed of nucleons, but at the same time to group the nucleons into cluster subsets in the theoretical description. The resonating group method (RGM), first proposed by Wheeler in 1937,[182] is the first attempt in this direction where the total wavefunction of the nucleus is the antisymmetrized product of the cluster and the inter-cluster wavefunctions.[183] The variables of a cluster wavefunction are the relative positions of the nucleons of that cluster, which somehow describe cluster's "internal" structure. On the other hand, each inter-cluster wavefunction describes the relative position of a pair of clusters using the relative position of the center of masses of the pair as the proper variable. Another similar and simpler approach, first proposed by Margenau,[184] and elaborated further by Hill and Wheeler,[185] is usually called Brink or Brink-Bloch model (See particularly Brink's contribution in Ref. 186).[172,186] In this model the basic wavefunction is the Slater determinant of the cluster wavefunctions, each localized in a certain point of the real-space;

based on the Hill-Wheeler method, these wavefunctions can be superposed using a weight function where the relative position of cluster pairs act as its variable. This elaborated form of Brink's wavefunction is transformable into corresponding RGM wavefunction revealing the tight connection of these two methods.[187,188] Brink's model has been used extensively for ab initio calculations on the cluster structures of many nuclei and one of the outcomes of these calculations is the nucleon densities.[189–195] The topographies of these densities are similar to the usual electronic densities promoting the idea that the topological analysis may reveal the underlying cluster structure in these densities. Although, to the best of author's knowledge, like the case of the Hadronic molecules discussed previously, no such study has been conducted yet. Surprisingly, some of these nucleon densities are easily reproducible also by the deformed harmonic oscillator model that is usually used as a simple model of the nuclear energy levels.[196,197] It is even more fascinating to perform the topological analysis on the unbiased nucleon densities, which are obtained without the assumption of the cluster structure *a priori* in the ab initio nuclear wavefunctions, derived directly from the "theory of everything". One of the simplest unbiased ab initio methods that starts from single nucleons, using a Slater determinant wavefunction composed of the one-nucleon wavefunctions, is the antisymmetrized molecular dynamics (AMD),[198,199] which recovers the clustered nucleon densities.[200] The derived clustered structures were further confirmed with more intricate ab initio studies that take various inter-nuclear correlations and even the relativistic effects into account.[201,202] All these open up a new domain for developing a novel methodology, the "quantum theory of clusters in nuclei" (QTCIN), to derive the cluster structure of nuclei from ab initio nuclear wavefunctions. Indeed, recent applications of the electron localization function (ELF),[203–205] to nuclei,[206–208] called nucleon localization function (NLF) in this new context,[209] have demonstrated the usefulness of concepts developed originally for the electronic matter in studying the nuclear matter. Finally, let us mention the intriguing possibility of the "condensation" of $\alpha$ clusters in the nuclei,[210–213] as a kind of the Bose-Einstein condensation,[214] which is currently an active field of research.[215–218] The present author has discussed briefly the possibility of the topological analysis of a bosonic system some time ago,[219] and clustered nuclei may be a real opportunity to apply this idea. All in all, the detailed origin of the nuclear clustering is yet considered an open problem,[220–222] and a version of the QTCIN, if developed properly, may help to clarify the very nature of this phenomenon from the analysis of the nuclear ab initio calculations.[223,224]

The last class considered herein is the exotic molecules, which will be categorized into four main subclasses in this contribution.[225–229] The first subclass includes "pure antimatter" which is derived by replacing all composing quantum particles of a molecule by corresponding antiparticles. The second/third subclass is the set of exotic molecules containing usually one but rarely more negatively/positively charged particles in addition to electrons and nuclei. The forth subclass is called hereafter "truly exotic species" where at least one of the main ingredients of usual molecules, i.e. electrons or nuclei, are totally absent from the system. Let us consider each subgroup briefly.

Pure antimatter is composed of the anti-particles; an antiparticle has exactly the same properties of its corresponding particle except from the sign of its electric charge.[9] The simplest and probably the most and best studied antimatter system is the antihydrogen atom composed of one positron, $e^+$,[9] the antiparticle of electron, and one antiproton, $p^-$.[230,231] Based on basic symmetries of the standard model of particle physics,[9] the antihydrogen atom is conceived as the exact "mirror image" of the hydrogen atom having the same properties and energy spectrum.[232] While predicted theoretically long time ago,[232] this antiatom has been "synthesized" only recently using accelerator beams,[233–236] and its optical spectrum has been studied intensively since then.[237–242] In principle nothing prevents the production of heavier antinuclei,[243,244] and indeed the antihelium nuclei have been produced recenetly.[245,246] However, gathering antinuclei and positrons for synthesizing antiatoms, and their subsequent trapping and spectroscopic studies, all pose real technological challenges. In fact, while antiatoms are as stable as atoms, since matter and antimatter annihilate upon coming together,[9] the storage and study of antiatoms, in a world made of atoms, require considerable efforts. Thus, it is not hard to imagine that synthesizing even the simplest antimatter molecules, e.g. the antimatter hydrogen molecule composed of two antihydrogen atoms, will be a real experimental achievement. On the other hand, if the basic symmetries of the standard model are universally correct, the world of antimatter is quite boring for a theorist since it is the exact mirror image of the matter world. As will be discussed in subsequent section, this makes the extension of the QTAIM to the antimolecules quite straightforward.

The second subclass includes exotic atoms where a heavy negatively charged particle is attached to an atom or a molecule. Examples are muon, $\mu^-$, $m_\mu \approx 207 m_e$, pion, $m_{pion} \approx 273 m_e$, kaon, $m_{kaon} \approx 966 m_e$ and antiproton; both pion and kaon are mesons while muon belongs to the lepton family and is a heavier congener of electron.[9] Some atomic members of the second subclass have been known for a long time which include pionic,[247] muonic,[248,249] Kaonic,[250–252] and antiprotonic,[253,254] atoms. None of these atoms are stable since not only pion and kaon are intrinsically unstable, with half-lifes less than a microsecond,[9] but also all these particles may participate in the "weak" or the strong nuclear interactions and finally being absorbed by the nucleus of the atom.[255–259] In fact, they are usually captured initially in orbitals with high atomic quantum numbers and then fall into lower energy orbitals, the lowest one with an orbital radius comparable to the size of nucleus, through the de-excitation process and upon emitting photons in X-ray wavelength.[260–264] This is easily understandable since from the basic hydrogen atom problem in quantum mechanics, it is well-known that the mean distance of the particle revolving around the nucleus is inversely proportional to its mass.[265] Consequently, the mean distances for the heavy negatively charged particles are hundreds of time smaller than the mean electron-nucleus distances. Accordingly, the spectroscopy of these exotic atoms may yield unique information about properties of nuclei as well as the forces operative within nuclei that are hard to be reached from other sources.[266–270] One may conclude that in the ground and probably some of the low-lying excited states the exotic atoms are not solely "Columbic" systems and the mentioned intricate interactions of the negatively charged particle and nucleus must be taken into account even in simplified quantum models of these atoms.[259,271] The same is true when the negatively charged particles are captured by molecules,[272] though in present contribution, for reasons to be disclosed in the next subsection, these intricacies are all ignored.

The third subclass includes all exotics where any one of the positively charged elementary particles is added to the usual atoms and molecules, though in practice, the positronic and muonic systems are the best known and most studied species from this subclass.[273–275] In fact, through various positron annihilation spectroscopies,[276–278] the positron emission tomography,[279,280] and the muon spin resonance spectroscopies ($\mu SR$),[281–291] a wealth of information is available on the positronic and muonic species particularly in condensed phases.[292–314] Of particular interest to chemists are muonic organic, organometallic, and biochemical molecules where a muonium atom, composed of an electron and a positive muon, $\mu^+$, bonds to a molecule making a radical species with an unpaired electron, which is studied through the $\mu SR$.[315–336] Even more interestingly, recent computational studies reveal that both $e^+$ and $\mu^+$ may participate directly in novel and unprecedented forms of bonding including the vibrational bonding,[337,338] and the one- and two-positron bonds.[48,339–343] Thus, our focus on the developments of the MC-QTAIM were on these species and more details will be disclosed in the subsequent subsection.

The subclass of the truly exotic species contains a heterogeneous set of systems where a package of particles through their attractive interactions make bound states, and the backbone is no longer a given usual atom or molecule. In this brief survey, we will cherry-pick only a handful of members of this set to disclose some of their interesting traits. Note that confirming the stability of a few-body quantum system, even when the interactions are solely the familiar Coulomb interactions, is not a trivial task theoretically and the stability/instability of some proposed species has been uncertain for decades.[344] Let us start from systems composed exclusively from electrons and positrons which were proposed theoretically as stable species by Wheeler in 1946.[345–347] The positronium atom, $Ps$, composed of an electron and a positron, is the prime example which was discovered experimentally by Deutsch in 1951,[348,349] and is one of the most studied systems of this subclass.[350–352] Since then the positronium negative ion, $Ps^-$, $Ps$ plus an electron,[353–355] and also $PsH$, $Ps$ plus a hydrogen atom,[356,357] were also discovered experimentally though based on our proposed classification scheme, $PsH$ belongs to the third subclass. Nevertheless, probably the most interesting discovery of this field was the production of what is called the "molecular positronium", $Ps_2$, composed of two electrons and two positrons.[355,358–360] Systems composed of electrons and positrons, as well as from heavier leptons, are "clean" Columbic systems.[361,362] However, this is not the case when the constituents of two-body exotic atoms are solely mesons or hadrons, e.g. pion and kaon,[363] where the strong interactions are also operative in addition to the Coulomb interactions.[364] Protonium atom, composed of $p^-$ and $p^+$, is an interesting example of these systems, which has been produced in accelerators and studied through the de-excitation process and the X-ray spectrum,[258,365–369] similar to the processes detailed previously for the pionic and kaonic atoms. Nevertheless, probably the most interesting examples, like the case of the third subclass, are those exotic molecular-like species where the particle responsible for "bonding" is not an electron. The first example is a Columbic three-body species composed of a $\mu^-$ and two of hydrogen isotopes, i.e. proton, deuterium and tritium, and is a practically a very compact diatomic molecule,[370] which has an important role also in the muon-catalyzed fusion.[371,372] In fact, this species is "isomorphic" to the hydrogen ion molecule where $\mu^-$ now plays the role of bonding agent instead of electron. The next example is a three-body system, which is not a purely Columbic system and is composed of two protons and a negatively charged kaon, where the latter now plays the role of bonding agent through both Coulomb and strong nuclear interactions.[373,374] All these examples and even more complex examples,[375–377] reveal that clustering and the intermediate level of organization in few- and many-body quantum systems are commonplace and it is desirable to describe this organization within the context of a unified scheme.

### II.C. AIM as conceived in the MC-QTAIM

Hopefully, the previous discussions convinced the skeptical reader that the extension of the QTAIM's paradigm beyond the purely electronic systems is due and the subject is worth pursuing. As emphasized in the introduction, from the three main discussed classes only a subclass of the exotic species, namely, the positronic and muonic systems have been considered computationally within the context of the MC-QTAIM.[40–48] Like the case of the QTAIM, the first step in the analysis is deciphering the inter-atomic boundaries and this is done through the local zero-flux equation of the gradient of the Gamma density: $\vec{\nabla}\Gamma^{(s)}(\vec{r}).\vec{n}(\vec{r})=0$.[33] The Gamma density is the mass-scaled sum of the one-particle densities of all quantum particles of the system: $\Gamma^{(s)}(\vec{r}) = \rho_1(\vec{r}) + \sum_{n=2}^{s}(m_1/m_n)\rho_n(\vec{r})$, where the subscript "1" is given to the lightest quantum particle (usually electron or positron). Each one-particle density is defined as follows: $\rho_n(\vec{r}) = N_n \int d\tau'_n \Psi^*\Psi$, where $\Psi$ is the wavefunction of a multi-component quantum system, while $m_n$ and $N_n$ are the mass and the number of the particles of $n-th$ subset, respectively, and $d\tau'_n$ implies summing over spin variables of all quantum particles and integrating over spatial coordinates of all quantum particles except one arbitrary particle belonging to the $n-th$ subset. The contribution of the one-particle density of each particle to the Gamma density is scaled according to its inverse mass relative to the lightest particle. The superscript $s$ is called the "cardinal number"; the QTAIM is a special case of the MC-QTAIM where $s=1$, while for the positronic and muonic species $s=2$ (this version may also be called the two-component QTAIM or the TC-QTAIM).[31] The purely antimatter molecules are also single-component cases, $s=1$, and the resulting local zero-flux equation, $\vec{\nabla}\rho_{positron}(\vec{r}).\vec{n}(\vec{r})=0$, is reminiscent of the local zero-flux equation of the QTAIM where $\rho_{positron}(\vec{r})$ is the counterpart of the electron density. In fact, this reveals the previously emphasized complete symmetry of the matter and the antimatter molecular structures and more generally the complete symmetry in matter and antimatter chemistries.

The original motivation for the introduction of the Gamma density was the extension of the regional virial theorem and the subsystem vibrational procedure of the QTAIM to the multi-component quantum systems within the context of the MC-QTAIM.[33] Nevertheless, the ultimate test of the effectiveness of the corresponding local zero-flux equation was its success in delineating reasonable inter-atomic boundaries in the positronic and muonic species.[40–48] Based on this success, an extended version of the BC is proposed for the two-component systems as follows:

*For an exotic molecular system containing $N_e$ electrons, with position vectors $\{\vec{r}_i\}$ and inter-electronic distances $\{r_{ij} = |\vec{r}_i - \vec{r}_j|\}$, and $q$ clamped nuclei each with atomic numbers $\{Z_\alpha\}$, with position vectors $\{\vec{R}_\alpha\}$ and inter-nuclear distances $\{R_{\alpha\beta} = |\vec{R}_\alpha - \vec{R}_\beta|\}$, and $N_p$ quantum particles, with a unit of positive charge and with a mass $m_p$ ( $m_p \geq m_e$ ), and with position vector $\{\vec{q}_k\}$ and inter-particle distances $\{q_{kl} = |\vec{q}_k - \vec{q}_l|\}$, described by the following Hamiltonian in atomic units:*

$$\hat{H} = \left(-\frac{1}{2}\right)\sum_i^{N_e} \nabla_i^2 + \left(-\frac{1}{2m_p}\right)\sum_k^{N_p} \nabla_k^2 + \sum_i^{N_e}\sum_{j>i}^{N_e} \frac{1}{r_{ij}} - \sum_i^{N_e}\sum_k^{N_p} \frac{1}{|\vec{r}_i - \vec{q}_k|} + \sum_k^{N_p}\sum_{l>k}^{N_p} \frac{1}{q_{kl}} + \hat{V}_{ext},$$

$$\hat{V}_{ext} = -\sum_i^{N_e}\sum_\alpha^Q \frac{Z_\alpha}{|\vec{R}_\alpha - \vec{r}_i|} + \sum_k^{N_p}\sum_\alpha^Q \frac{Z_\alpha}{|\vec{R}_\alpha - \vec{q}_k|}$$

*There is always a critical mass, $m_c$, and a critical distance between each pair of nuclei, $\{R_{\alpha\beta}^c\}$, that above these critical values $m_p > m_c$, $\vec{\nabla}\Gamma^{(2)}_{ground-state}\left(\vec{r}; \{R_{\alpha\beta} > R_{\alpha\beta}^c\}\right)$ contains* *just $P = Q + N_p$ numbers of (3, -3) CPs where $Q$ of them are located at the position of the clamped nuclei whereas for $m_p < m_c$, $\vec{\nabla}\Gamma^{(2)}_{ground-state}\left(\vec{r}; \{R_{\alpha\beta} > R^c_{\alpha\beta}\}\right)$ contains just $Q$ numbers of (3, -3) CPs at the location of the clamped nuclei.*

This conjecture is hereafter called the extended Bader's conjecture (EBC) and apart from the previously raised points regarding the BC, this extended version needs some extra clarifications.

In the case of the positronic and muonic species: $\rho_1(\vec{r}) = \rho_e(\vec{r})$ and $\rho_2(\vec{r}) = \rho_P(\vec{r})$, thus: $\Gamma^{(2)}(\vec{r}) = \rho_e(\vec{r}) + (m_e/m_p)\rho_p(\vec{r})$. $\Gamma^{(2)}(\vec{r})$ is a "combined" density,[42] and this means that the appearance of (3, -3) CPs is a convoluted act of the two one-particle densities and the relative masses of the constituent particles, all contributing to the topography of $\Gamma^{(2)}(\vec{r})$. Apart from their direct contribution to $\Gamma^{(2)}(\vec{r})$, the masses are present in the Hamiltonian and so they influence indirectly the topography of the one-particle densities through shaping the multi-component wavefunction. Deriving the exact mass-dependence of the one-particle densities is not an easy task analytically. By the way, computational studies reveal that for $m_p \approx m_e$, e.g. the positronic species, $\rho_p(\vec{r})$ is a very diffuse and flat function thus contributes marginally to the main topographical features of $\Gamma^{(2)}(\vec{r})$. In this limit, the number of (3, -3) CPs of $\Gamma^{(2)}(\vec{r})$ and $\rho_e(\vec{r})$ are both equal to $Q$ and all are located at the clamped nuclei that means the positrons are unable to shape their own atomic basins. Indeed, a wealth of independent studies on the positronic densities,[339,340,378–395] apart from ours,[41,48] conform to this picture. In other words, the positron is contained within one (or rarely two) atomic basins shaped by the clamped nuclei and effectively, $\Gamma^{(2)}(\vec{r})$ only reveals the electron clustering.[41,48] The other extreme is: $m_p \gg m_e$, captured mathematically by the limit: $m_e/m_p \to 0$, is easy to be studied analytically if one assumes that each heavy quantum particle acts like a harmonic oscillator.[32] In this limit $\rho_p(\vec{r})$ is a very localized delta-like function and it is possible to demonstrate analytically that: $\lim_{m_e/m_p \to 0} \Gamma^{(2)}(\vec{r}) \to \rho(\vec{r})$,[32] where $\rho(\vec{r})$ is the usual electron density used within the context of the QTAIM (not to be confused with $\rho_e(\vec{r})$). Practically, this limit yields the clamped proton in the adiabatic view and the number of (3, -3) CPs of $\Gamma^{(2)}(\vec{r})$ is equal to $Q + N_P$ since there are now effectively $Q + N_P$ clamped nuclei, and the EBC "reduces" to the BC. Once again, in this limit also $\Gamma^{(2)}(\vec{r})$ captures effectively only the electron clustering surrounding the clamped nuclei and the localized quantum particles. One may conclude that somewhere between: $m_p \approx m_e$ and $m_p \gg m_e$, the positive quantum particle acquires the capacity to shape its own atomic basin. The exact numerical value of this critical mass is not fixed and changes among various species. The phenomenon has been termed the "topological structural transformation" since the $Q$-atomic basin system transforms into $Q + N_P$-atomic basin species.[44] However, since we know that proton is almost always capable of forming its own atomic basin in molecules, then: $m_{positorn} < m_c < m_{proton}$, and interestingly, muon's mass is in this range. Our computational studies on the muonic molecules,[43,45,46] neglecting some special cases,[44] revealed that the positive muon is indeed capable of forming its own atomic basin, thus for many molecular species: $m_c < m_\mu$.

It is remarkable that apart from some very simple positronic species with one or no nuclei, which are hard to be classified either as atom or molecule,[396–411] based on majority of both experimental,[412–415] and theoretical literature,[378–395] molecular structure are not seriously altered upon the addition of positron. Whereas, for the muonic molecules a wealth of experimental and theoretical studies lead to the picture that $\mu^+$ is the "light radioisotope" of hydrogen,[416–429] and forms its own atomic basin. All these conform well to the predictions derived from the EBC and since the only constraint of the conjecture is $m_p \geq m_e$, the EBC must be applicable to every real or hypothetical particle as far as the dominant interactions are the Coulomb interactions. Probably the next natural choice is a molecule containing the heavier tau lepton, $m_{tau} \approx 3477 m_e$, as the last known member of the lepton famliy.[9] However, because of its much smaller lifetime compared to muons, $\sim 10^{-13} s$ vs. $\sim 10^{-6} s$,[9] there are no unambiguous experimentally detected "tauonic" atoms and molecules. Nevertheless, if observed in future, the EBC predicts concretely that tau, because of its large mass, which is almost equal to that of deuterium, must be capable to form its own atomic basin in yet hypothetical tauonic molecules and act as a "heavy radioisotope" of hydrogen. Let us stress that it seems reasonable to extend the EBC to encompass multi-component systems containing $s$-type of quantum particles, replacing $\Gamma^{(s)}(\vec{r})$ with $\Gamma^{(2)}(\vec{r})$ in the conjecture. There are indeed computational and theoretical evidence supporting this extension,[43] however we prefer to relegate this possibility before further studies.

The curious reader may wonder why the EBC does not cover the case of the exotic molecules containing the negatively charged quantum particles. In fact, it is feasible to propose a more extended version of the EBC to encompass these species as well, however, as articulated below, the resulting extension is of no chemical significance. Let's start from considering the situation with a computationally studied example, namely, a five-body system composed of a $\mu^-$, a point $\alpha$ particle, a proton, and two electrons.[43] Our ab initio calculations clearly revealed that this system is effectively a diatomic species composed of a very compact cluster of $\alpha + \mu^-$, which bonds to proton through a two-electron bond. This "internal clusterization" is easily verifiable geometrically since the mean $(\alpha, \mu^-)$ distance is more than one hundred times smaller than the mean $(\alpha, p^+)$ and $(\mu^-, p^+)$ distances. Also, the virial theorem, dictating the balance between the expectation values of the kinetic and potential energies, $\langle \hat{T} \rangle = (-1/2)\langle \hat{V} \rangle$,[430] retains for the cluster revealing the "decoupling" of the cluster's dynamics from rest of the system. Accordingly, one may replace the cluster with a hypothetical point particle with a unit of positive charge (the net charge of the cluster), and a mass of $m_{cluster} = m_\alpha + m_\mu$, and redo the ab initio calculations on this four-body system. The resulting electronic structure is virtually indistinguishable from that of the original five-body system and is quite similar to a normal hydrogen molecule.[43] In a subsequent highly accurate ab initio study Frolov considered the internal clusterization in $ab\mu^- e_2$ and $abc\mu^- e_2$ systems where $a, b, c$ stand for any of the three hydrogen isotopes and $e$ for electron.[431] Various computed geometrical mean values in these systems revealed that in the case of $ab\mu^- e_2$ species, each system is composed of a very compact cluster of $ab\mu^-$ (*vide supra*) where the two electrons revolve around the cluster; practically resembling a hydrogen anion or a hydride. In the case of $abc\mu^- e_2$ species, the same study revealed the very compact cluster of $ab\mu^-$ is bonded to the lighter isotope $c$ through the two-electron bond, similar to a hydrogen molecule. Interestingly, since the compact $ab\mu^-$ is itself a diatomic species, like the Russian dolls, $abc\mu^- e_2$ structure is effectively a compact diatomic system within a more spatially extended diatomic structure.[431] Since $\mu^-$ is the lightest negatively charged particle apart from electron, in the case of other negatively charged particles even more compact clusters are formed and all these are in line with the previously discussed traits of the negatively charged particle containing within exotic atoms. Based on these examples, the main role of a negatively charged particle in molecules is "screening" the nuclear charge and diminishing the effective atomic number of its centering nucleus by one unit, as seen by electrons. In the case of $\mu^-$, Reyes and coworkers have derived the very localized negative muon's one-particle density around the clamped nucleus by ab initio calculations.[432–434] They properly resemble this extreme nuclear charge screening as a kind of "alchemical transmutation" and the same is true for the role of other negatively charged particles in the exotic atoms and molecules. In such cases instead of the deducing inter-atomic boundaries through the local zero-flux equation of the MC-QTAIM, one may instead use the local zero-flux equation of the QTAIM. The proper effective Hamiltonian for molecules with charge-screened clamped nuclei is the usual electronic Hamiltonian where the atomic number of each nucleus, with a negatively charged particle revolving around, changes from $Z$ to $Z-1$. In next step, one may derive the electron density from the ab initio computed effective ground-state electronic wavefunction and then incorporate it into the local zero-flux equation of the QTAIM. What if one insists to use the local zero-flux equation of the MC-QTAIM in such situations using $\Gamma^{(2)}(\vec{r}) = \rho_e(\vec{r}) + (m_e/m_N)\rho_N(\vec{r})$ (the subscript $N$ stands for negative)? Evidently, since the negatively charged particle is localized around the clamped nucleus, its one-particle density, $\rho_N(\vec{r})$, is a delta-like function centered at the nucleus where the one-particle electron density, $\rho_e(\vec{r})$, is also maximum. Thus, $\rho_N(\vec{r})$ only affects $\Gamma^{(2)}(\vec{r})$ at and immediately around the nucleus by elevating the height of the Gamma density without changing its topography. In fact, $\rho_e(\vec{r})$ solely dictates the topography of $\Gamma^{(2)}(\vec{r})$ while the electron density derived from the previously-mentioned effective model accurately reproduces $\rho_e(\vec{r})$. To sum up, the number of atomic basins does not vary before and after the addition of the negatively charged particles to the molecular system and only the properties of the atomic basin containing the particle vary because of the variation in the nuclear charge. The only remaining technical difficulty is the possible effect of the extra terms in the Hamiltonian, describing the absorption of the negatively charged particles into the nucleus,[271,435] on the Gamma density, which must clarified in future studies.

Finally, let us briefly mention the case of the truly exotic systems where because of some technical obstacles, the effectiveness of the current version of the EBC for these species is less clear than for the other above-considered exotic molecules. The main problem in dealing with the truly exotic systems is the fact that the mass ratio of the constituent particles is usually not as small as that of the mass ratio of electrons to nuclei. Thus, clamping certain particles in these systems and making an adiabatic approximation from the outset for ab initio calculations, similar to the Born-Oppenheimer (BO) procedure for usual molecules,[436–438] is not legitimate *per se*. The only safe way of dealing with these systems is to treat all constituent particles as quantum particles, i.e. each having both kinetic and potential energy operators in the Hamiltonian, and to solve Schrödinger's equation in a fully non-BO scheme.[439–441] Apart from the fact that non-BO calculations are more computationally demanding than those within the adiabatic approximation, there are also serious "qualitative" differences between the properties and symmetries of wavefunctions derived within and beyond the BO paradigm.[442–457] Herein, we will not discuss these differences and leave this subject to a future study where all details will be reviewed (Shahbazian, under preparation). However, let us just stress that in deriving both BC and EBC the presence of the "laboratory-fixed" framework is a necessary condition *per se* which is absent beyond the adiabatic approximation. Thus, future developments of new methodologies of deriving AIM structure from molecular non-BO wavefunctions, either usual or exotic, must be based on the "body-fixed" quantities and concepts, as is also evident from the recent non-BO studies.[455–457]

## III. AIM properties beyond purely electronic systems

### III.A. AIM properties as conceived within the QTAIM

Within the context of the QTAIM, each atomic basin, $\Omega$, receives its share of molecular properties, i.e. usually molecular expectation values, $\langle \hat{A} \rangle$, as follows: $\langle \hat{A} \rangle = \sum_{k=1}^{Q} A(\Omega_k)$ (hereafter it is assumed that the number of basins in a molecule is equal to the number of clamped nuclei).[14,26,27] $A(\Omega_k) = \int_{\Omega_k} d\vec{r}\, A(\vec{r})$, is the contribution of $k-th$ basin from property $\hat{A}$ and, $A(\vec{r}) = \int d\tau'\, \text{Re}\left[\Psi^* \hat{A} \Psi\right]$ is the property density.[14,458] For the case of the one-electron operators, $\hat{M} = \sum_{i=1}^{N} \hat{m}_i$, e.g. the kinetic energy, the property density, $M(\vec{r}) = \hat{m}(\vec{r}) \rho^{(1)}(\vec{r}', \vec{r})\big|_{\vec{r}=\vec{r}'}$, is expressed using the reduced spinless first-order density matrix (1-RDM): $\rho^{(1)}(\vec{r}', \vec{r}) = N \sum_{spins} \int d\vec{r}_2 ... \int d\vec{r}_N \Psi^*(\vec{r}_1', \vec{r}_2,...) \Psi(\vec{r}_1, \vec{r}_2,...)$.[14,459] In the original formulation of the QTAIM,[14] for the two-electron operators, $\hat{G} = \sum_{i=1}^{N} \sum_{j>i}^{N} \hat{g}_{ij}$, e.g. the electron-electron repulsion term, the property density, $G(\vec{r}) = (1/2) \int d\vec{r}_2 \hat{g}(\vec{r}_1, \vec{r}_2) \rho^{(2)}(\vec{r}_1, \vec{r}_2)$, is expressed by the diagonal part of the reduced spinless second-order density matrix (2-RDM): $\rho^{(2)}(\vec{r}_1, \vec{r}_2) = N(N-1) \sum_{spins} \int d\vec{r}_3 ... \int d\vec{r}_N \Psi^* \Psi$.[14,459] However, as demonstrated by Popelier, Pendás and coworkers in their proposed "interacting quantum atoms" energy partitioning scheme,[460–464] it is much more informative to partition the expectation values of the two-electron properties into intra- and inter-basin contributions: $\langle \hat{G} \rangle = \sum_{k=1}^{Q} G(\Omega_k) + \sum_{k=1}^{Q} \sum_{l>k}^{Q} G(\Omega_k, \Omega_l)$. Assuming that the operators do not include spatial derivatives, these contributions are also expressible employing the diagonal part of the 2-RDM:

$$G(\Omega_k) = (1/2) \int_{\Omega_k} d\vec{r}_1 \int_{\Omega_k} d\vec{r}_2 \hat{g}(\vec{r}_1, \vec{r}_2) \rho^{(2)}(\vec{r}_1, \vec{r}_2), \qquad G(\Omega_k, \Omega_l) = \int_{\Omega_k} d\vec{r}_1 \int_{\Omega_l} d\vec{r}_2 \hat{g}(\vec{r}_1, \vec{r}_2) \rho^{(2)}(\vec{r}_1, \vec{r}_2).$$

Generally, the inter-basin contribution is a quantitative measure of the interaction and/or communication between the two atomic basins $\Omega_k$ and $\Omega_l$. In one sense, the whole QTAIM analysis aims for "chemical interpretation" of the computed basin and inter-basin properties as well as their variations in various molecules or the conformers of the same moleucle.[14,26,27] Interestingly, when it came to the electron population in atomic basins, Bader bypassed his original density-based formalism and introduced the concept of the electron number fluctuation.[14,49,50] This is the basis of definition of the so-called electron delocalization index used to quantify the degree of covalency and the bond orders (*vide infra*).[465,466] Let us consider some details of this procedure briefly.

Incorporating $\hat{A} = N\hat{1}$ into the above formalism yields the mean basin electron population: $N(\Omega_k) = \int_{\Omega_k} d\vec{r}\, \rho(\vec{r})$, $N = \sum_{k=1}^{Q} N(\Omega_k)$.[14,467] Nevertheless, with some theoretical arguments, Bader conceived atomic basins as "open subsystem" with "non-vanishing fluctuations" of properties (See page 171 in Ref. 14). Pursuing this line of thought and in order to introduce electron number fluctuation, the basin "electron number distribution" was introduced: $P_m(\Omega_k) = \binom{N}{m} \int_{\Omega_k} d\vec{r}_1 \ldots \int_{\Omega_k} d\vec{r}_m \int_{R^3-\Omega_k} d\vec{r}_{m+1} \ldots \int_{R^3-\Omega_k} d\vec{r}_N \sum_{spins} \Psi^*\Psi$, and because of the normalization of wavefunction one arrives at: $\sum_{m=0}^{N} P_m(\Omega_k) = 1$.[14,49,50] Each $P_m(\Omega_k)$ is the probability of observing an "image" of molecule in which there is an $m$-electron cluster, $0 \le m \le N$, in the $k-th$ basin while the rest of $N-m$ electrons are in $R^3 - \Omega_k$. It is straightforward to demonstrate that the electron population is also directly derivable as the mean value of this distribution: $N(\Omega_k) = \sum_{m=1}^{N} m P_m(\Omega_k)$.[49] At the next step, the electron fluctuation is introduced as the variance of the distribution as follows: $\Lambda(\Omega_k, N) = N^2(\Omega_k) - [N(\Omega_K)]^2 = \sum_{m=1}^{N} m^2 P_m(\Omega_k) - \left[\sum_{m=1}^{N} m P_m(\Omega_k)\right]^2$.[49] Alternatively, it is straightforward to derive $\Lambda(\Omega_k, N)$ directly from the 2-RDM: $\Lambda(\Omega_k, N) = \int_{\Omega_k} d\vec{r}_1 \int_{\Omega_k} d\vec{r}_2 \rho^{(2)}(\vec{r}_1, \vec{r}_2) + \int_{\Omega_k} d\vec{r}\, \rho(\vec{r}) - \left[\int_{\Omega_k} d\vec{r}\, \rho(\vec{r})\right]^2$.[14,49,50] Generally: $\Lambda(\Omega_k, N) > 0$, but for the whole system as a closed system: $\Lambda(R^3, N) = 0$, which is in line with the fact that while the electrons fluctuate between basins, the total number of electrons is a conserved quantity. This line of reasoning has never been extended to include other properties apart from electron number thus the mentioned "non-vanishing fluctuations" of properties has never been quantified. Based on this background, we may introduce the general idea of property fluctuations of atomic basins for the one-electron properties and then considering some of its immediate ramifications.

In quantum mechanics the variance of a property, corresponding to a quantum state, is computed as follows: $\Lambda(R^3, M) = \langle \hat{M}^2 \rangle - \langle \hat{M} \rangle^2$, and it is only null for the constants of motion, i.e. the properties commuting with the Hamiltonian of molecule.[468] Assuming $R^3$ to be divided into an atomic basin, $\Omega_1$, and the complement region, $\Omega_2 = R^3 - \Omega_1$, it is straightforward to decompose the total variance into the basin variances and the inter-basin covariance:

$$\Lambda(R^3, M) = \Lambda(\Omega_1, M) + \Lambda(\Omega_2, M) + 2Cov_M(\Omega_1, \Omega_2)$$

$$\Lambda(\Omega_k, M) = M^2(\Omega_k) - [M(\Omega_k)]^2, \quad Cov_M(\Omega_1, \Omega_2) = M(\Omega_1, \Omega_2) - M(\Omega_1)M(\Omega_2)$$

Where:

$$M^2(\Omega_k) = \int_{\Omega_k} d\vec{r} \; \text{Re}\left[\hat{m}^2(\vec{r})\rho^{(1)}(\vec{r}',\vec{r})\right]\bigg|_{\vec{r}'=\vec{r}}$$

$$+ \int_{\Omega_k} d\vec{r}_1 \int_{\Omega_k} d\vec{r}_2 \; \text{Re}\left[\hat{m}_1(\vec{r}_1)\hat{m}_2(\vec{r}_2)\rho^{(2)}(\vec{r}_1',\vec{r}_2',\vec{r}_1,\vec{r}_2)\right]\bigg|_{\vec{r}_i'=\vec{r}_i}$$

$$M(\Omega_1, \Omega_2) = \int_{\Omega_1} d\vec{r}_1 \int_{\Omega_2} d\vec{r}_2 \; \text{Re}\left[\hat{m}_1(\vec{r}_1)\hat{m}_2(\vec{r}_2)\rho^{(2)}(\vec{r}_1',\vec{r}_2',\vec{r}_1,\vec{r}_2)\right]\bigg|_{\vec{r}_i'=\vec{r}_i}$$

$$\rho^{(2)}(\vec{r}_1',\vec{r}_2',\vec{r}_1,\vec{r}_2) = N(N-1) \sum_{spins} \int d\vec{r}_3 ... \int d\vec{r}_N \Psi^*(\vec{r}_1',\vec{r}_2',\vec{r}_3...) \Psi(\vec{r}_1,\vec{r}_2,\vec{r}_3...)$$

$\rho^{(2)}(\vec{r}_1', \vec{r}_2', \vec{r}_1, \vec{r}_2)$ in these expressions is the general non-diagonal form of the 2-RDM. Also, it is easy to check that as a matter of consistency, one recovers the previously derived expression for $\Lambda(\Omega_k, N)$ by incorporating $\hat{m} = 1$ in these equations. For constants of motion: $\Lambda(\Omega_1, M) + \Lambda(\Omega_2, M) = -2Cov_M(\Omega, \Omega')$, which asserts that the total system is closed in regard to the property $M$. All these results are easily extendable to an exhaustive partitioning of $R^3$ into $Q$ atomic basins, $\bigcup_k \Omega_k = R^3$, finally yielding:

$$\Lambda(R^3, M) = \sum_{k=1}^{Q} \Lambda(\Omega_k, M) + 2\sum_{k=1}^{Q}\sum_{l>k}^{Q} Cov_M(\Omega_k, \Omega_l).$$ In the case of the number of electrons, the index of electron number delocalization is introduced as follows: $\delta_N(\Omega_k, \Omega_l) = 2|Cov_N(\Omega_k, \Omega_l)|$,[465] which is a measure of the inter-basin electron sharing. In the same spirit, one may propose the index of property delocalization: $\delta_M(\Omega_k, \Omega_l) = 2|Cov_M(\Omega_k, \Omega_l)|$, which contains the original electron number delocalization as a special case. However, in contrast to the electron number distribution, there is no analogous property distribution that could be used to compute the mean, $M(\Omega_k)$, and variance, $\Lambda(\Omega_k, M)$, simultionsly. In order to compute $M(\Omega_k)$, the following distribution may be introduced: $M_m(\Omega_k) = \binom{N}{m} \int_{\Omega_k} d\vec{r}_1 ... \int_{\Omega_k} d\vec{r}_m \int_{R^3 - \Omega_k} d\vec{r}_{m+1} ... \int_{R^3 - \Omega_k} d\vec{r}_N \sum_{spins} \text{Re}[\Psi^* \hat{m}(\vec{r}_1) \Psi]$. This is a generalized form of $\{P_m(\Omega_k)\}$, but in contrast to $\{P_m(\Omega_k)\}$, $\{M_m(\Omega_k)\}$ has the physical dimension of the property $\hat{M}$. Since $\langle \hat{M} \rangle / N = \sum_{m=0}^{N} M_m(\Omega_k)$, therefore each $M_m(\Omega_k)$ is a contribution of $\langle \hat{M} \rangle / N$ attributed to an "image" in which $m$-electrons are in $\Omega_k$ while the

rest of $N-m$ electrons are in $R^3 - \Omega_k$. It is also straightforward to demonstrate that:

$$M(\Omega_k) = \sum_{m=1}^{N} m M_m(\Omega_k) = \int_{\Omega_k} d\vec{r} M(\vec{r}),$$ thus recovering the basin properties originally deduced

from the formulation discussed previously. However, as stressed, $M^2(\Omega_k) \neq \sum_{m=1}^{N} m^2 M_m(\Omega_k)$,

and $\Lambda(\Omega_k, M)$ is "not" the variance of $\{M_m(\Omega_k)\}$ distribution. This result reveals the special trait of the electron number distribution since in this case: $\hat{m} = \hat{m}^2 = \hat{1}$, while in general for an interacting system of particles even when $\hat{M}$ is a constant of motion, each $\hat{m}_i$ is not a constant

of motion and thus there is no simple relationship between $M_m(\Omega_k)$ and $P_m(\Omega_k)$. Let us

finally stress that the role of $M(\Omega_k, \Omega_l)$ for the one-electron properties is similar to that of

$G(\Omega_k, \Omega_l)$ for the two-electron properties; both yield the inter-basin contributions although,

the latter appears when computing the basin properties, while the former only appears when

computing the basin property fluctuations. In fact, $\delta_M(\Omega_k, \Omega_l)$ is gauging the amount of

"openness" of atomic basins relative to various properties so this index is the quantitative manifestation of Bader's original idea of the property fluctuation.

### III.B. AIM properties as conceived within the MC-QTAIM

The same idea of property partitioning is also applicable within the context of the MC-

QTAIM to the $s$-component molecular systems, containing $\sum_{n=1}^{s} N_n = N_t$ quantum particles,

and partitioned into $P$ atomic basins through the zero-flux equation of $\Gamma^{(s)}(\vec{r})$ and the EBC.

In the case of the one-particle properties, $\hat{M} = \sum_{n=1}^{s} \sum_{i=1}^{N_n} \hat{m}_{n,i}$, the expectation values are partitioned as follows: $\langle \hat{M} \rangle = \sum_{k=1}^{P} \tilde{M}(\Omega_k)$, where $\tilde{M}(\Omega_k) = \sum_{n=1}^{s} M_n(\Omega_k)$ is the total basin contribution and $M_n(\Omega_k) = \int_{\Omega_k} d\vec{r}\, M_n(\vec{r})$ is the contribution of $n-th$ subset of particles to the $k-th$ basin.[33,37]

The corresponding property density is as follows: $M_n(\vec{r}) = \hat{m}_n(\vec{r}) \rho_n^{(1)}(\vec{r}',\vec{r})|_{\vec{r}=\vec{r}'}$, in which $\rho_n^{(1)}(\vec{r}',\vec{r}) = N_n \int d\tau_n' \Psi^*(...,\vec{r}_{n,1},...,\vec{r}_{n,i}',...\vec{r}_{n,N_n},...) \Psi(...,\vec{r}_{n,1},...,\vec{r}_{n,i},...\vec{r}_{n,N_n},...)$ is the generalized form of the 1-RDM for the $n-th$ subset of particles where its diagonal part yields the previously introduced one-particle density, $\rho_n(\vec{r})$.[32–34] In the case of the two-particle operators there are two types of operators namely, those acting between members of the same component and those operative between members of two different components, $\hat{G} = \sum_{n=1}^{s}\left(\sum_{i=1}^{N_n}\sum_{j>i}^{N_n} \hat{g}_{n,ij}\right) + \sum_{n=1}^{s}\sum_{m>n}^{s}\left(\sum_{i=1}^{N_n}\sum_{j=1}^{N_m} \hat{g}_{nm,ij}\right)$. Employing the rational of partitioning the expectation value into intra- and inter-basin contributions one arrives at: $\langle \hat{G} \rangle = \sum_{k=1}^{P} \tilde{G}(\Omega_k) + \sum_{k=1}^{P}\sum_{l>k}^{P} \tilde{G}(\Omega_k,\Omega_l)$, where $\tilde{G}(\Omega_k) = \sum_{n=1}^{s}\left[G_n(\Omega_k) + \sum_{m>n}^{s} G_{nm}(\Omega_k)\right]$ and $\tilde{G}(\Omega_k,\Omega_l) = \sum_{n=1}^{s}\left[G_n(\Omega_k,\Omega_l) + \sum_{m>n}^{s} G_{nm}(\Omega_k,\Omega_l)\right]$. Assuming that the two-particle operators do not contain spatial derivatives, the terms in brackets are computed as follows: $G_n(\Omega_k) = (1/2)\int_{\Omega_k} d\vec{r}_1 \int_{\Omega_k} d\vec{r}_2\, \hat{g}_{nn}(\vec{r}_1,\vec{r}_2)\rho_n^{(2)}(\vec{r}_1,\vec{r}_2)$, $G_{nm}(\Omega_k) = \int_{\Omega_k} d\vec{r}_n \int_{\Omega_k} d\vec{r}_m\, \hat{g}_{nm}(\vec{r}_n,\vec{r}_m)\rho_{nm}^{(2)}(\vec{r}_n,\vec{r}_m)$, $G_n(\Omega_k,\Omega_l) = \int_{\Omega_k} d\vec{r}_1 \int_{\Omega_l} d\vec{r}_2\, \hat{g}_{nn}(\vec{r}_1,\vec{r}_2)\rho_n^{(2)}(\vec{r}_1,\vec{r}_2)$, $G_{nm}(\Omega_k,\Omega_l) = \int_{\Omega_k} d\vec{r}_n \int_{\Omega_l} d\vec{r}_m\, \hat{g}_{nm}(\vec{r}_n,\vec{r}_m)\rho_{nm}^{(2)}(\vec{r}_n,\vec{r}_m)$, in which $\rho_n^{(2)}(\vec{r}_1,\vec{r}_2) = N_n(N_n-1)\int d\tau_n'' \Psi^*\Psi$, and $\rho_{nm}^{(2)}(\vec{r}_n,\vec{r}_m) = N_n N_m \int d\tau_n''' \Psi^*\Psi$, which is the pair density. In these integrations $d\tau_n''$ implies summing over spin variables of all quantum particles

and integrating over spatial coordinates of all quantum particles except two arbitrary particles, denoted herein as particles "1" and "2", belonging to the $n-th$ subset. While, $d\tau_n'''$ implies summing over spin variables of all quantum particles and integrating over spatial coordinates of all quantum particles except two arbitrary particles, one belongs to the $n-th$ and the other to the $m-th$ subsets. Also, $\rho_n^{(2)}(\vec{r}_1,\vec{r}_2)$ is the generalized form of the 2-RDM for the $n-th$ subset of particles whereas $\rho_{nm}^{(2)}(\vec{r}_n,\vec{r}_m)$, which emerges because of the "distinguishability" of the two particles belonging to two different components, is a novel pair density foreign to the purely electronic systems.[32–34] In addition, it is straightforward to demonstrate that for the single-component systems, all these equations reduce to the partitioning schemes of the QTAIM discussed in the previous subsection. Let us now consider the case of the particle fluctuation within context of the MC-QTAIM.

Incorporating $\hat{m}=\hat{1}$ into the one-particle property density yields the basin populations for each type of particles: $N_n(\Omega_k) = \int_{\Omega_k} d\vec{r}\, \rho_n(\vec{r})$, $N_t = \sum_{k=1}^{P}\sum_{n=1}^{s} N_n(\Omega_k)$.[31,34] In order to introduce the particle number fluctuation, the basin particle number distribution for each subset is defined as follows: $P_m^n(\Omega_k) = \binom{N_n}{m}\int_{\Omega_k} d\vec{r}_{n,1}...\int_{\Omega_k} d\vec{r}_{n,m} \int_{R^3-\Omega_k} d\vec{r}_{n,m+1}...\int_{R^3-\Omega_k} d\vec{r}_{n,N} \int d\omega_n \Psi^*\Psi$, where $d\omega_n$ implies summing over spin variables of all quantum particles and integrating over spatial coordinates of all quantum particles except the $n-th$ subset while the normalization condition of the multi-component wavefunctions implies: $\sum_{m=0}^{N_n} P_m^n(\Omega_k) = 1$.[34] Each $P_m^n(\Omega_k)$ is the probability of observing an "image" of molecule in which there is an $m$-particle cluster from the $n-th$ subset, $0 \leq m \leq N_n$, in the $k-th$ basin while the rest of $N_n - m$ particles are in $R^3 - \Omega_k$. The basin populations of each type of particles are derived as the mean value of each distribution: $N_n(\Omega_k) = \sum_{m=1}^{N_n} m P_m^n(\Omega_k)$, while the particle fluctuation is introduced as the variance of these distributions: $\Lambda_n(\Omega_k, N_n) = N_n^2(\Omega_k) - [N_n(\Omega_K)]^2 = \sum_{m=1}^{N_n} m^2 P_m^n(\Omega_k) - \left[\sum_{m=1}^{N_n} m P_m^n(\Omega_k)\right]^2$.

The variance may alternatively be deduced from the generalized form of the 2-RDM as follows: $\Lambda_n(\Omega_k, N) = \int_{\Omega_k} d\vec{r}_1 \int_{\Omega_k} d\vec{r}_2 \rho_n^{(2)}(\vec{r}_1, \vec{r}_2) + N_n(\Omega_k) - [N_n(\Omega_k)]^2$.[34] For the particles effectively confined into a single basin, the corresponding variance is null: $\Lambda_n(\Omega_k, N_n) \approx 0$,[34,36] but in general: $\Lambda_n(\Omega_k, N_n) > 0$, while: $\Lambda_n(R^3, N_n) = 0$, where the latter is the result of particle number conservation. It is straightforward to demonstrate that:

$$\sum_{k=1}^{P} \Lambda_n(\Omega_k, N_n) + 2 \sum_{k=1}^{P} \sum_{l>k}^{P} Cov_n^N(\Omega_k, \Omega_l) = 0,$$

where the inter-basin covariances, $Cov_n^N(\Omega_k, \Omega_l) = \int_{\Omega_k} d\vec{r}_1 \int_{\Omega_l} d\vec{r}_2 \rho_n^{(2)}(\vec{r}_1, \vec{r}_2) - N_n(\Omega_k) N_n(\Omega_l)$, are used to introduce the delocalization index for each subset of particles: $\delta_n^N(\Omega_k, \Omega_l) = 2|Cov_n^N(\Omega_k, \Omega_l)|$.[34,36] Apart from the above-mentioned distributions, a novel type of joint distribution may also be introduced for the multi-component systems, which has not been considered previously in the MC-QTAIM literature. The joint basin particle number distribution for a couple of subsets is defined as follows:

$$P_{mp}^{nq}(\Omega_k) = \binom{N_n}{m}\binom{N_q}{p} \int_{\Omega_k} d\vec{r}_{n,1} \ldots \int_{\Omega_k} d\vec{r}_{n,m} \int_{R^3-\Omega_k} d\vec{r}_{n,m+1} \ldots \int_{R^3-\Omega_k} d\vec{r}_{n,N_n} \int_{\Omega_k} d\vec{r}_{q,1} \ldots \int_{\Omega_k} d\vec{r}_{q,p} \int_{R^3-\Omega_k} d\vec{r}_{q,p+1} \ldots \int_{R^3-\Omega_k} d\vec{r}_{q,N_q} \int d\omega_{nq} \Psi^* \Psi,$$

where $d\omega_{np}$ implies summing over spin variables of all quantum particles and integrating over spatial coordinates of all quantum particles except the spatial coordinates of the $n-th$ and the $q-th$ subsets. Each $P_{mp}^{nq}(\Omega_k)$ is the joint probability of observing an "image" of a molecule in which there is an $m$-particle cluster from the $n-th$ subset, $0 \leq m \leq N_n$, and a $p$-particle cluster from the $q-th$ subset, $0 \leq p \leq N_q$, in the $k-th$ basin while the rest of $N_n - m$ and $N_q - p$ particles are in $R^3 - \Omega_k$. The previously introduced basin particle number distributions are derivable from the joint distribution as follows: $P_m^n(\Omega_k) = \sum_{p=1}^{N_q} P_{mp}^{nq}(\Omega_k)$, $P_p^q(\Omega_k) = \sum_{m=1}^{N_n} P_{mp}^{nq}(\Omega_k)$, while the normalization of the wavefunction implies that:

$$\sum_{m=1}^{N_n}\sum_{p=1}^{N_q} P_{mp}^{nq}(\Omega_k) = 1.$$ The covariance of the joint distribution is formally defined as follows:

$$Cov_{nq}^N(\Omega_k) = \sum_{m=1}^{N_n}\sum_{p=1}^{N_q}(m - N_n(\Omega_k))(p - N_q(\Omega_k))P_{mp}^{nq}(\Omega_k),^{469}$$ which after some mathematical manipulations may alternatively be expressed using the pair density as follows:

$$Cov_{nq}^N(\Omega_k) = \int_{\Omega_k} d\vec{r}_n \int_{\Omega_k} d\vec{r}_q \rho_{nq}^{(2)}(\vec{r}_n, \vec{r}_q) - \int_{\Omega_k} d\vec{r}\, \rho_n(\vec{r}) \int_{\Omega_k} d\vec{r}\, \rho_q(\vec{r}).$$ It is straightforward to demonstrate that: $\sum_{k=1}^P Cov_{nq}^N(\Omega_k) + 2\sum_{k=1}^P \sum_{l>k}^P \overline{Cov}_{nq}^N(\Omega_k, \Omega_l) = 0$, where the novel inter-basin covariances are defined as follows:

$$\overline{Cov}_{nq}^N(\Omega_k, \Omega_l) = (1/2)\left[\int_{\Omega_k} d\vec{r}_n \int_{\Omega_l} d\vec{r}_q \rho_{nq}^{(2)}(\vec{r}_n, \vec{r}_q) - N_n(\Omega_k)N_q(\Omega_l) + \int_{\Omega_l} d\vec{r}_n \int_{\Omega_k} d\vec{r}_q \rho_{nq}^{(2)}(\vec{r}_n, \vec{r}_q) - N_q(\Omega_k)N_n(\Omega_l)\right].$$

Whether $Cov_{nq}^N(\Omega_k)$ and $\overline{Cov}_{nq}^N(\Omega_k, \Omega_l)$ may found proper chemical interpretations like the case of $Cov_n^N(\Omega_k, \Omega_l)$ is an open problem that needs further theoretical and computational investigations. At this stage of development, we may introduce the basin property fluctuations.

Like the case of the QTAIM discussed previously, our starting point is the variance of a one-particle property for the total molecular system: $\Lambda(R^3, M) = \langle \hat{M}^2 \rangle - \langle \hat{M} \rangle^2$. Assuming $R^3$ to be divided into $Q$ atomic basins, $\bigcup_k \Omega_k = R^3$, it is straightforward to decompose the total variance into various basin and inter-basin covariances:

$$\Lambda(R^3, M) = \sum_{k=1}^{P}\left[\sum_{n=1}^{s}\Lambda_n(\Omega_k, M)\right] + 2\sum_{k=1}^{P}\left[\sum_{n=1}^{s}\sum_{q>n}^{s} Cov_{nq}^M(\Omega_k)\right]$$

$$+ 2\sum_{k=1}^{P}\sum_{l>k}^{P}\left[\sum_{n=1}^{s} Cov_n^M(\Omega_k, \Omega_l)\right] + 4\sum_{k=1}^{P}\sum_{l>k}^{P}\left[\sum_{n=1}^{s}\sum_{q>n}^{s} \overline{Cov}_{nq}^M(\Omega_k, \Omega_l)\right]$$

where:

$$\Lambda_n(\Omega_k, M) = M_n^2(\Omega_k) - [M_n(\Omega_k)]^2, \quad Cov_n^M(\Omega_k, \Omega_l) = M_n(\Omega_k, \Omega_l) - M_n(\Omega_k)M_n(\Omega_l)$$

$$M_n^2(\Omega_k) = \int_{\Omega_k} d\vec{r}\; \text{Re}\left[\hat{m}_n^2(\vec{r})\rho_n^{(1)}(\vec{r}',\vec{r})\right]\bigg|_{\vec{r}'=\vec{r}}$$

$$+ \int_{\Omega_k} d\vec{r}_1 \int_{\Omega_k} d\vec{r}_2\; \text{Re}\left[\hat{m}_n(\vec{r}_1)\hat{m}_n(\vec{r}_2)\rho_n^{(2)}(\vec{r}_1',\vec{r}_2',\vec{r}_1,\vec{r}_2)\right]\bigg|_{\vec{r}_i'=\vec{r}_i}$$

$$M_n(\Omega_k, \Omega_l) = \int_{\Omega_k} d\vec{r}_1 \int_{\Omega_l} d\vec{r}_2\; \text{Re}\left[\hat{m}_n(\vec{r}_1)\hat{m}_n(\vec{r}_2)\rho_n^{(2)}(\vec{r}_1',\vec{r}_2',\vec{r}_1,\vec{r}_2)\right]\bigg|_{\vec{r}_i'=\vec{r}_i}$$

$$\rho_n^{(2)}(\vec{r}_1',\vec{r}_2',\vec{r}_1,\vec{r}_2) = N_n(N_n-1)\int d\tau_n''\; \Psi^*(\vec{r}_{n,1}',\vec{r}_{n,2}',\vec{r}_{n,3}...)\Psi(\vec{r}_{n,1},\vec{r}_{n,2},\vec{r}_{n,3}...)$$

and:

$$Cov_{nq}^M(\Omega_k) = M_{nq}(\Omega_k) - M_n(\Omega_k)M_q(\Omega_k),$$

$$\overline{Cov}_{nq}^M(\Omega_k, \Omega_l) = (1/2)\left[M_{nq}(\Omega_k, \Omega_l) - M_n(\Omega_k)M_q(\Omega_l)\right] + (1/2)\left[M_{nq}(\Omega_l, \Omega_k) - M_n(\Omega_l)M_q(\Omega_k)\right]$$

$$M_{nq}(\Omega_k) = \int_{\Omega_k} d\vec{r}_n \int_{\Omega_k} d\vec{r}_q\; \text{Re}\left[\hat{m}_n(\vec{r}_n)\hat{m}_q(\vec{r}_q)\rho_{nq}^{(2)}(\vec{r}_n,\vec{r}_q)\right]$$

$$M_{nq}(\Omega_k, \Omega_l) = \int_{\Omega_k} d\vec{r}_n \int_{\Omega_l} d\vec{r}_q \, \text{Re}\left[\hat{m}_n(\vec{r}_n)\hat{m}_q(\vec{r}_q)\rho_{nq}^{(2)}(\vec{r}_n,\vec{r}_q)\right]$$

$$M_{nq}(\Omega_l, \Omega_k) = \int_{\Omega_l} d\vec{r}_n \int_{\Omega_k} d\vec{r}_q \, \text{Re}\left[\hat{m}_n(\vec{r}_n)\hat{m}_q(\vec{r}_q)\rho_{nq}^{(2)}(\vec{r}_n,\vec{r}_q)\right]$$

$\rho_n^{(2)}(\vec{r}_1',\vec{r}_2',\vec{r}_1,\vec{r}_2)$ in these expressions is the non-diagonal form of the previously introduced generalized 2-RDM for the $n-th$ subset of particles. Interestingly, the contributions from joint distributions, i.e. $Cov_{nq}^M(\Omega_k)$ and $\overline{Cov}_{nq}^M(\Omega_k,\Omega_l)$, emerge automatically from the partitioning of the total variance. As far as the property $\hat{M}$ is not a constant of motion, the total variance is non-zero, $\Lambda(R^3,M) > 0$, however, even if it is a constant of motion, $\Lambda(R^3,M) = 0$, it not possible to separate the one-particle and joint contributions into two groups and the sum of contributions of each group is not separately equal to zero. This stems from the fact that particles from different subsets may exchange properties thus in contrast to the number of particles, there is no "local" conservation law for the properties of each subset. Accordingly, it is not possible to deduce a conservation law for the sum of the variances and covariances associated to the $n-th$ subset of particles nor the joint distribution of the $n-th$, and the $p-th$, subsets of particles. By the way, the sum of the all contributions of particle number fluctuations is recovered if one incorporates $\hat{m} = 1$ into the property fluctuation expression. One may propose: $\delta_n^M(\Omega_k,\Omega_l) = 2|Cov_n^M(\Omega_k,\Omega_l)|$, as the index of property delocalization for the $n-th$ subset that contains the previously defined particle number delocalization as a special case. However, as detailed previously, it is important to realize that $\delta_n^M(\Omega_k,\Omega_l)$, in contrast to $\delta_M(\Omega_k,\Omega_l)$, is "not" the sole contribution of the inter-basin property fluctuation and the role of $\overline{Cov}_{nq}^M(\Omega_k,\Omega_l)$ must be also taken into account. Like the case of $Cov_{nq}^N(\Omega_k)$ and $\overline{Cov}_{nq}^{N}(\Omega_k,\Omega_l)$, future theoretical and computational studies may shed some light on possible chemical interpretations of $Cov_{nq}^{M}(\Omega_k)$ and $\overline{Cov}_{nq}^{M}$.

## IV. Conclusion

The primary goal of an AIM-based partitioning scheme is to propose an explanation for the origin of stability of a molecular system in comparison to sum of its constituent atoms or compared to other isomers/conformers of the molecule. This explanation is "coarse-grained" by its nature since from the viewpoint of the "theory of everything", the ultimate source of the stability of all quantum few- and many-body systems are the stabilizing interactions between the constituent elementary particles. The effectiveness of such coarse-grained explanation critically depends on its ability to "locate" the origin of stability to one or at most to a small number of interactions between atomic basins. In other words, the spatially "non-local" explanation of stability within context of the "theory of everything", which spreads throughout the molecule, is replaced by a spatially "localized" explanation. Whether this is feasible for a special molecular system or a group of systems in a comparative study is a matter of computational considerations and is beyond the theoretical analysis of the foundations of a partitioning methodology. This is also the case for the introduced property delocalization index but taking the fact that deducing the inter-basin contribution of the two-electron interactions had a huge success in unrevealing the origin of the local stabilizing interactions,[463] one may reasonably expect that the proposed index to be informative as well.

Also, the idea of clusterization in few-body and many-body quantum systems need further studies beyond the previous reports. Let us stress that a system with well-separated clusters of particles may be considered in an "adiabatic" paradigm as discussed by Frolov.[431] The Hamiltonian of a clustered system may be divided into the cluster Hamiltonians, each containing the internal dynamics of the cluster plus an extra potential energy term simulating the effect of other clusters. After solving Schrödinger's equation for each cluster, similar to the usual BO procedure,[436,437] the dynamical couplings between clusters may also be taken into account as "non-adiabatic" corrections. This tacitly implies that there is a link between the "degree of clusterization" at the ground state and the energy spectrum of the system, which needs further scrutiny in future theoretical studies. Accordingly, beyond calculating the average inter-particle distances, which is practical only for systems composed of distinguishable particles, it is desirable to seek for more universal "indices of clusterization". While the BC and EBC are examples in this regard, in cases where the clusterization has multiple layers, and clusters are formed independently at different spatial scales,[43,431] novel procedures and concepts must be introduced. The most desirable procedure is to infer the clusterization from the Hamiltonian itself, i.e. the properties of quantum particles and their modes of interaction, rather than the deriving them from analyzing the wavefunctions or from the computed expectation values.

The MC-QTAIM analysis of the exotic molecules started almost a decade ago in our laboratory, and arguably the most important achievement in the initial phase of developments was the demonstration that $\mu^+$ has the capacity to form its own atomic basin.[43,45,46] This adds a new type of atom in a molecule, an exotic one, to the known AIM, but, the recent discovery of the positron bond,[339] and its subsequent MC-QTAIM analysis,[48] revealed the capacity of the MC-QTAIM analysis in tracing and quantifying the "exotic bonds" as well. The fact that agents other than electrons may act as the bonding glue in the exotic species, has the massage that the usual chemical bonds, discovered in the purely electronic systems with all their ramifications, are just a single subclass among a large class of yet to be identified exotic bonds.

All these support the viewpoint that concepts of AIM and bonds, although originally invented in chemistry, are also applicable to few-body quantum systems that bear no apparent resemblance to the usual molecules. A large amount of theoretical developments and computational applications remains to be done in this area as well as in the case of the Hadronic and nuclear molecules mentioned previously.

## Acknowledgments

The author is grateful to Mohammad Goli for his constructive comments on a previous draft.